\documentclass[a4paper,fleqn,usenatbib]{mnras}
\usepackage{amssymb,amsmath,graphicx,psfrag,mathtools}
\def\app#1#2{%
	\mathrel{%
		\setbox0=\hbox{$#1\approx$}%
		\setbox2=\hbox{%
			\rlap{\hbox{$#1\propto$}}%
			\lower1.1\ht0\box0%
			}%
			\raise0.25\ht2\box2%
			}%
			}

\usepackage[T1]{fontenc} %MNRAS
\usepackage{ae,aecompl} %MNRAS
\usepackage{url}
\usepackage{revsymb}
\hypersetup{draft} %TO AVOID ERROR: \PDFENDLINK ENDED UP IN DIFFERENT
\title[A 1091 s White Dwarf Pulsar]{GLEAM-X J162759.5$-$523504.3 as a White Dwarf Pulsar}
\author[J. I. Katz]{
	J. I. Katz,$^{1}$\thanks{E-mail katz@wuphys.wustl.edu} %MNRAS
\\
$^{1}$Department of Physics and McDonnell Center for the Space Sciences,
Washington University, St. Louis, Mo. 63130 USA %MNRAS
}
\date{Accepted XXX.  Received YYY; in original form ZZZ} %MNRAS
\pubyear{2022} %MNRAS
\date{\today}
\begin{document} %MNRAS
%\psfrag{theta}{$\theta$}  Works only on COMPLETE strings
\label{firstpage} %MNRAS
\pagerange{\pageref{firstpage}--\pageref{lastpage}} %MNRAS
\maketitle %MNRAS
\begin{abstract}
	The low frequency radio source GLEAM-X J162759.5$-$523504.3 emits
	pulsed coherent polarized emission like that of radio pulsars.  Yet
	its period of 18.18 min (1091 s) is hundreds of times longer than
	those of confirmed radio pulsars, and if it is a neutron star its
	mean radiated power exceeds the upper limit on its spin-down power
	by more than an order of magnitude.  This may be explained if it is
	a white dwarf pulsar, with a moment of inertia several orders of
	magnitude greater than those of neutron stars.  The Lorentz factor
	of the emitting charge bunches may be bounded from below by the
	widths of the temporal substructure of their radiation.  If the
	emission is curvature radiation, the radius of curvature may be
	estimated from the Lorentz factor and the frequencies of emission;
	it is consistent with a white dwarf's inner magnetosphere but not a
	neutron star's.
\end{abstract}
 %MNRAS
\section{Introduction}
``Pulsar'' has had many meanings.  Originally it referred to periodic
coherent radio emission from rotating magnetic neutron stars, that were soon
discovered to emit periodic but incoherent visible light, X-rays and
gamma-rays.  It has sometimes been broadened to mean any periodic emission
from a rotating object, including thermal X-rays in ``X-Ray Pulsars''.

Here I will consider a pulsar to be an object that emits periodic pulses of
{\it coherent\/} radio radiation, whether or not that is accompanied by 
emission at other frequencies.  Coherent emission requires plasma processes
that create charge ``bunches''.  Astronomical objects that emit coherently
are naturally grouped together, even if their parameters differ
quantitatively.

The original pulsars were rotating neutron stars, effectively in vacuum.
Most of them are single; the minority in binary systems are separated from
their companions by distances much greater than their radii of light
cylinders $R = c/\Omega$, where $\Omega$ is their angular frequency.
Such binary pulsars radiate as if they were single, although they may
produce winds that interact with their companions, and mass-losing
companions may turn the pulsar into an accreting X-ray source that does not
emit coherent radio radiation.
\section{A White Dwarf Pulsar}
Since the early days of pulsar astronomy there has been speculation that
a rotating magnetic white dwarf might show pulsar-like activity.  The words
``white dwarf pulsar'' appear in over 100 abstracts, most of which discuss
incoherent periodic emission, usually of visible light.  One binary star, AR
Sco, with periodic incoherent radio and visible emission at the beat
frequency between a white dwarf's rotation and the orbital frequency, has
been proposed as a white dwarf pulsar \citep{MGH16}.  However, coherent
radio emission has not been reported and its companion is well within the
white dwarf's radius of light cylinder, suggesting that the physical
processes involved differ from those of traditional radio pulsars.  Models
involving interaction between the two stars have been developed
\citep{GZH16,K17}.

The recently discovered \citep{HW22} periodic radio transient GLEAM-X
J162759.5$-$523504.3 is a candidate for the first true white dwarf pulsar.
It has a period of 18.18 minutes (1091 s) and its pulses show low
frequency (72--215 MHz) emission with a brightness temperature $\sim 10^{16}
\,$K implying coherent emission.  It has no binary companion with which to
interact.  It thus meets the criteria of a classical pulsar, although its
period is hundreds of times longer than any of theirs.

GLEAM-X J162759.5$-$523504.3 differs from all known classical pulsars in one
important way.  The measured upper bound on the rate of change of its period
implies, using the usual estimate of a neutron star's moment of inertia of
$\sim 10^{45}\,$g-cm$^2$, an upper bound on its spin-down power of $1.2
\times 10^{28}\,$erg/s \citep{HW22}.  This is exceeded by its instantaneous
pulse luminosity (in the band of observation, surely an underestimate) of up
to $4 \times 10^{31}\,$ergs/s.  Averaging over the pulse duty factor of
about 1\% (this is equivalent to assuming the radiation is emitted into a
fraction of $4\pi$ steradians equal to the measured duty factor) indicates a
mean radiated luminosity $\sim 4 \times 10^{29}\,$ergs/s.  This may be an
overestimate because we may be favorably situated near the peak of its beam
pattern (observational selection makes that likely), but the mean radiated
power still exceeds the spindown power by more than an order of magnitude.
This is physically impossible for a rotation-powered object.  No classical
pulsar emits more than about 1\% of its spindown power as coherent radio
emission \citep{SZM14}.

The obvious resolution of these two difficulties---the anomalously long
rotation period and the insufficient spindown power---is that the object
is a white dwarf rather than a neutron star.  White dwarves have moments
of inertia $\sim 10^{50}\,$g-cm$^2$, about five orders of magnitude greater
than that of a neutron star, increasing the estimated spindown power to
$\sim 10^{33}\,$ergs/s, sufficient to power the radio emission with
plausible ($\ll 1$) efficiency.

Because of their large moments of inertia, white dwarves rotate more slowly
than neutron stars; GLEAM-X J162759.5$-$523504.3 would be an unusually
fast-rotating white dwarf without accretional spin-up, but not the fastest
known.  For example, SDSS J125230.93$-$023417.72 has a period of 317 s
\citep{RHV20}, WD 1832$+$089 has a period of 353 s \citep{PDB20}, RE
J0317$-$853 has a period of 725 s \citep{BJOD95,FVW97} and ZTF
J190132$+$145808.7 has a period of 6.94 minutes \citep{CBF21}.  The
photometric variability is likely attributable to strong magnetic fields,
nonuniform over the surface.  ZTF J190132$+$145808.7 has magnetic fields in
the 600--900 MG range, while the magnetic field of RE J0317$-$853 is several
hundred MG and SDSS J125230.93$-$023417.72 has a starspot with a field of 5
MG.  No measurement of the magnetic field of GLEAM-X J162759.5$-$523504.3
exists, but the similarity among these objects hints that it might also be
strongly magnetic, which would be consistent with its being a pulsar.
\section{Lorentz Factors}
In order for radiation from a rotating emitter with period $P$ to have
temporal structure of full width at half-maximum (FWHM) $\Delta t$ it must
be beamed into an angle
\begin{equation}
	\theta \approx {2 \pi \Delta t \over P}.
\end{equation}
In GLEAM-X J162759.5$-$523504.3 temporal structure as fine as $\Delta t
\approx 0.5\,$s has been observed (Fig.~3 of \citet{HW22}), implying
$\theta \lesssim 3 \times 10^{-3}\,$radian.  If this radiation is produced
by accelerated relativistic charges, their Lorentz factor $\gamma$ must
satisfy \citep{RL79}
\begin{equation}
	\label{gamma}
	\gamma \gtrapprox 0.9/\theta \approx 300.
\end{equation}
This inequality is an approximate equality for a narrowly collimatedi
particle beam.  Charges whose velocity vectors are spread by angles larger
than the reciprocal of their Lorentz factor produce a radiation pattern set
by their angular dispersion rather than by their Lorentz factor, making
Eq.~\ref{gamma} an inequality.

Curvature radiation is expected for charges in strong magnetic fields, in
which their transverse motion rapidly decays by synchrotron radiation, or is
not excited because nonresonant electric fields perpendicular to strong
magnetic fields are not effective accelerators.  If curvature radiation is
the emission process then the radius of curvature $\rho$ may be estimated
from the expression for the peak of the spectrum of emission by a
relativistic particle \citep{J99}
\begin{equation}
	\nu_{peak} \approx 0.3 \gamma^3 {c \over \rho},
\end{equation}
or
\begin{equation}
	\rho \approx 0.3 {\gamma^3 c \over \nu_{peak}} \gtrsim 1.5 \times
	10^9\,\text{cm},
\end{equation}
where Eq.~\ref{gamma} has been used to set a bound on $\gamma$, and
$\nu_{peak} \approx 200\,$MHz is the middle of the frequency band of
Fig.~3 of \citet{HW22}, from which $\theta$ and $\gamma$ have been bounded.
This is consistent with radiation in a white dwarf magnetosphere, but not in
a neutron star magnetosphere.  It could be reconciled with a neutron star
magnetosphere only if radiation extended to frequencies of hundreds of GHz,
which would exacerbate the (already forbidding) energetic problems with
neutron star models.
\section{Discussion}
The long period and emitted power of GLEAM-X J162759.5$-$523504.3 both point
to its nature as a white dwarf pulsar, the first such object identified
whose physics and radiation mechanism could resemble those of classical
radio pulsars.  This hypothesis is supported by the Lorentz factor inferred
from the temporal substructure of at least one of its pulses, that implies
a large radius of curvature of the magnetic field lines if the emission is
curvature radiation.

The possible similarity between GLEAM-X J162759.5$-$523504.3 and strongly
magnetic, fast-rotating, white dwarves suggests that the former object, a
coherent radio emitter, might be a promising target for optical observations
to test the prediction made here that it is a white dwarf, although at its
estimated distance of about 1 kpc it might be forbiddingly faint.  If it
were bright enough, optical observations could also determine its magnetic
field, spectroscopically or polarimetrically.  The fast-rotating, strongly
magnetized, white dwarves would be promising targets for low frequency radio
observations to determine if any of them are white dwarf pulsars.

The model discussed here differs from previous models.  \citet{LM22} predict
that a hot, luminous subdwarf will be found at the location of GLEAM-X
J162759.5$-$523504.3, while \citet{ES22} predict no visible object there, a
small $\dot P$, a $\sim 0.1\,$s spin periodicity in addition to the long
observed period, and likely a luminous pulsar wind nebula.
\section*{Data Availability}
This theoretical study did not generate any new data.

\label{lastpage}

\begin{thebibliography}{99}
	\bibitem[\protect\citeauthoryear{Barstow, Jordan, O'Donoghue {\it
		et al.\/}}{1995}]{BJOD95} Barstow, M. A., Jordan, S.,
		O'Donoghue, D. {\it et al.\/} 1995 \mnras\ 277, 971.
	\bibitem[\protect\citeauthoryear{Caiazzo, Burdge, Fuller {\it et
		al.\/}}{2021}]{CBF21} Caiazzo, I., Burdge, K. B., Fuller, J.
		{\it et al.\/} 2021 Nature 595, 39.
	\bibitem[\protect\citeauthoryear{Ek\c{s}i \& \c{S}a\c{s}maz}{2022}]
		{ES22} Ek\c{s}i, K. Y. \& \c{S}a\c{s}maz, S. 2022 \mnras\
		submitted arXiv:2202.05160.
	\bibitem[\protect\citeauthoryear{Ferrario, Vennes, Wickramasinghe
		{\it et al.\/}}{1997}]{FVW97} Ferrario, L., Vennes, S.,
		Wickramasinghe, D. T. {\it et al.\/} 1997 \mnras\ 292, 205.
	\bibitem[\protect\citeauthoryear{Geng, Zhang, Huang {\it et al.\/}}
		{2016}]{GZH16} Geng, J.-J., Zhang, B., Huang, Y. F. {\it et
		al.\/} 2016 \apjl\ 831, L10.
	\bibitem[\protect\citeauthoryear{Hurley-Walker, Zhang, Bahramian
		{\it et al.\/}}{2022}]{HW22} Hurley-Walker, N., Zhang, X.,
		Bahramian, A. {\it et al.\/} 2022 Nature 601, 526.
	\bibitem[\protect\citeauthoryear{Jackson}{1999}]{J99} Jackson, J. D.
		1999 {\it Classical Electrodynamics\/} (3rd ed.) Chichester:
		Wiley.
	\bibitem[\protect\citeauthoryear{Katz}{2017}]{K17} Katz, J. I. 2017
		\apj\ 835, 150.
	\bibitem[\protect\citeauthoryear{Loeb \& Maoz}{2022}]{LM22} Loeb, A.
		\& Maoz, D. 2022 Res. Notes AAS in press arXiv:2202.04949.
	\bibitem[\protect\citeauthoryear{Marsh, G\"{a}nsicke, H\"{u}mmerich
		{\it et al.\/}}{2016}]{MGH16} Marsh, T. R., G\"{a}nsicke, B.
		T., H\"{u}mmerich, S. {\it et al.\/} 2016 Nature 537, 374.
	\bibitem[\protect\citeauthoryear{Pshirkov, Dodin, Belinski {\it et
		al.\/}}{2020}]{PDB20} Pshirkov, M. S., Dodin, A. V.,
		Belinski, A. A. {\it et al.\/} 2020 \mnras\ 499, L21.
	\bibitem[\protect\citeauthoryear{Reding, Hermes, Vanderbosch {\it
		et al.\/}}{2020}]{RHV20} Reding, J. S., Hermes, J. J.,
		Vanderbosch, Z. {\it et al.\/} 2020 \apj\ 894, 19.
	\bibitem[\protect\citeauthoryear{Rybicki \& Lightman}{1979}]{RL79}
		Rybicki, G. B. \& Lightman, A. P. 1979 {\it Radiative
		Processes in Astrophysics\/} (New York: Wiley).
	\bibitem[\protect\citeauthoryear{Szary, Zhang, Melikidze {\it et
		al.\/}}{2014}]{SZM14} Szary, A., Zhang, B., Melikidze, G. I.
		{\it et al.\/} 2014 \apj\ 784, 59.
\end{thebibliography}
\end{document}